# QCD phase transition with two flavors of Wilson quarks using a RG improved action[1]

K. Kanaya

*Center for Computational Physics, Institute of Physics,*
*University of Tsukuba, Ibaraki 305, Japan*

## Abstract

The finite temperature QCD phase transition is studied on the lattice with degenerate two flavors of Wilson quarks. Motivated by reported strange behaviors with the standard action on lattices with the temporal extention $N_t = 4$ and 6, a renormalization group improved gauge action is applied. On an $N_t = 4$ lattice, the strange behaviors observed with the standard action are removed with our improved action. The finite temperature transition is continuous in the chiral limit and it becomes quite smooth in all observables we studied when we increase the quark mass by increasing $\beta$ along the crossover line.

---

# 1 Introduction and motivations

Understanding the nature of the finite temperature transition in QCD with two degenerate light quarks ($N_F = 2$) is a very important step towards the clarification of the QCD transition in the real world. For $N_F = 2$ QCD, Wilczek and Rajagopal have developed an argument that, if the transition in the chiral limit ("chiral transition") is of second order, then it will belong to the same univesality class as the three dimensional O(4) Heisenberg model [1]. This gives us several useful scaling relations near the chiral transition that we can use to discriminate the nature of the chiral transition. If the chiral transition is of second order, we expect that it turns into an analytic crossover at finite quark mass.

From numerical studies using staggered quarks [2] no signs of discontinuities are found for $m_q \geq 0.0125$ on $N_t = 4$ and 6 lattices, where $N_t$ is the temporal lattice size. Furthermore, recent scaling studies [3] show that behaviors of physical quantities are consistent with the predicted O(4) scaling near the chiral transition, suggesting that the chiral transition is of second order. However, because the action for $N_F = 2$ staggered quarks is not local, further study with Wilson quarks is desired.

The phase diagram for $N_F = 2$ QCD with Wilson quarks is summarized in Fig. 1 collecting recent data of various groups (see references in [4]). As in the case of staggered quarks, the chiral transition at $\beta = \beta_{ct}$ is shown to be continuous both on $N_t = 4$ and 6 lattices [5, 6]. However, a detailed study by MILC collaboration [7] show that on $N_t = 4$ lattices the transition becomes once very strong at $K \simeq 0.19$ when we decrease $K$ along the finite temperature transition/crossover $K_t$, and it becomes weaker again when we further decrease $K$. Fig. 2 shows $m_\pi^2$ at $\beta = 5.0$ obtained by us and MILC collaboration. We find that $m_\pi^2$ changes its behavior sharply at $K_t$. Other quantities, such as Polyakov loop and plaquette, also show a rapid change of behaviors at $K_t$. Although no discontinuities are observed on $N_t = 4$ lattices, the transition does not look like a crossover here. On an $N_t = 6$ lattice [8]

MILC group even found first order signals at several intermediate values of $K$ ($K = 0.17$, $0.18$ and $0.19$). Consulting the phase diagram Fig. 1 more closely, we, however, note that these points of strong $K_t$ are just the points where $K_t$ lines get once very close to $K_c$ line due to the rather sharp bend of $K_c$ at $\beta \simeq 5$ caused by the cross-over phenomenon between the weak and strong coupling regions of QCD. Therefore, it seems plausible that the strong $K_t$ is a lattice artifact caused by this unusual relation of $K_t$ and $K_c$ lines.

Another lattice artifact we encounter with Wilson quarks using the standard action is the strange behavior of quark mass $m_q$ at $\beta \lesssim 5.3$ in the deconfined phase [9, 7]. $m_q$ for Wilson quark is defined through an axial Ward identity [10]. Numerical studies for $\beta = 5.85$ in quenched QCD [11] and $\beta = 5.5$ in $N_F = 2$ QCD [12] show that (i) at zero temperature $m_\pi$ vanishes at $K$ which is almost identical to that where $m_q = 0$, and (ii) the value of $m_q$ does not depend on $N_t$ and the phase. At $\beta \lesssim 5.3$, however, the second property is lost. $m_q$ in the deconfined phase now shows dependence on $N_t$ and deviate from the value in the confining phase (Fig. 2) suggesting sizable $O(a)$ corrections to the axial Ward identity in the deconfined phase.

A naive way out of these lattice artifacts is to increase $N_t$ so that we have the chiral transition in the weak coupling region, $\beta_{ct} \gtrsim 5.5$. However, our previous study on a $18^3 \times 24$ lattice suggests that this requires $N_t > 18$ [6].

This motivated us to study these issues using improved actions. We apply a renormalization group (RG) improved action proposed by Iwasaki about ten years ago [13] which is reviewed in the next section. In section 3, results of numerical simulation performed on $8^3 \times 4$ and $8^4$ lattices are summarized. We find that the unsatisfactory features of the standard action discussed above are removed at $\beta$ down to the chiral transition point for $N_t = 4$.

# 2  Improved action

In the scaling region, a lattice action describes the continuum properties at sufficiently large distances in lattice units. For these distances, the symmetries expected in the continuum limt, such as the rotation symmetry, will recover. At short distances comparable with the lattice spacing $a$, however, lattice actions generally fail to reproduce continuum properties even in the scaling region. This makes MC simulations near the continuum limit very hard. On the other hand, lattice action has freedom to introduce less local terms without affecting the continuum limit. For example, lattice gauge action may contain rectangles, chairs, etc. besides the plaquettes adopted in the standard action: from loops up to length 6,

$$S_{gauge} = \beta \{ c_0 \sum \Box + c_1 \sum \Box\Box \; c_2 \sum \text{⌐┐} + c_3 \sum \text{▢} \},$$

where $\beta = 6/g^2$ and $c_0 + 8c_1 + 16c_2 + 8c_3 = 1$. Although the long distance behaviors are not affected by these additional terms, short distance properties are modified. Therefore, by adjusting these additional coupling parameters, the short distance lattice artifacts may be reduced. Such actions are called as improved actions.

Several strategies have been proposed to find a set of coupling parameters which improves the theory efficiently. Here, we apply a strategy based on a renormalization group (RG) transformation [13].

Consider a block transformation of scale factor 2. Repeated application of a block transformation defines a chain of effective actions (trajectory) approaching an IR fixed point $P_0$ in the infinite-dimensional coupling parameter space (Fig. 3). The hypersurface where the correlation length diverges is called as "critical surface". We suppose a fixed point $P_\infty$ on the critical surface. The trajectory connecting $P_\infty$ and $P_0$ is called as "renormalized trajectory" (RT). Actions on RT are "perfect" (completely improved) actions [14] in the sense that continuum behavior is realized from the distance of $1a$ (up to distortions depending on the choice of the block transformation) because RT is connected with the continuum theory $P_\infty$. When the initial action $S^{(0)}$ is in the scaling region that is close to the critical surface, the trajectory first flows along the critical surface and then leaves the critical surface near $P_\infty$ towards $P_0$ approaching RT gradually. The number of block transformations required to get sufficiently close to RT is nothing but $\log_2$ of the minimum distance to obtain continuum behaviors with $S^{(0)}$.

Hasenfratz and Niedermayer obtained a functional equation for $P_\infty$ in several asymptotically free theories [14] and proposed to use it (approximated on a finite dimensional parameter space) as an approximate perfect action.

Here, we are not aiming at a perfect action. Instead, we are trying to take advantage of the extended parameter space to accelerate the approach to RT. For example, consider an action $S_{imp}$ in the extended parameter space and suppose that the additional parameters in $S_{imp}$ are adjusted so that the action gets sufficiently close to RT after, say, 2 block transformations. This means that, when we simulate the system using $S_{imp}$, a separation of $4a$ is enough to get the continuum properties. In summary, our strategy is: (i) restrict ourselves to a small dimensional coupling parameter space consisting only with interactions which are easy to be implemented in numerical simulations, and (ii) find coupling parameters which minimizes the distance to RT after *a few* block transformations.

Iwasaki applied this program to the SU(3) gauge theory [13]. Because the theory is an asymptotically free theory, correlation functions close to the critical surface can be computed by perturbation theory: for Wilson loops $W(C) = 1 - g^2 F(C) + O(g^4)$, the first coefficient $F(C)$ can be easily computed. Choosing the block transformation sufficiently simple, blocked correlation functions can also be computed: $W^{(I)}(C) = 1 - g^2 F^{(I)}(C) + O(g^4)$ for $I$-th blocked lattice, even in the limit of RT ($I = \infty$). Iwasaki studied the 4 parameter gauge action consisting loops up to length 6. To the lowest order perturbation theory $c_2$ and $c_3$ appear only in the combination $c_2 + c_3$.

Using a simple block averaging for the block transformation:
$$A_\mu^{(I+1)}(n') = \frac{2}{2^d} \sum_{n \in n'} A_\mu^{(I)}(n)$$
where $A_\mu^{(I)}(n)$ is the vector potential on $I$-th blocked lattice $\{n\}$ and $n'$ is a hypercube, and by defining the distance to RT after $I$-blockings by
$$R^{(I)} = \sqrt{\sum_C (\frac{F^{(I)}(C) - F^{(\infty)}(C)}{F^{(\infty)}(C)})^2 / \sum_C 1}$$
with $\sum_C$ over loops up to length 6, Iwasaki found that both $R^{(1)}$ and $R^{(2)}$ form very narrow valley with rather flat bottom in the space $(c_1, c_2+c_3)$: For $c_1 \lesssim 0$ and $c_2 + c_3 \lesssim 0$, the valleys are given by $1.531 c_1 + c_2 + c_3 = -0.5061$ and $1.761 c_1 + c_2 + c_3 = -0.5116$ for $R^{(1)}$ and $R^{(2)}$, respectively, where the true minima locate at about $c_1 = -0.27$ for both cases. Smallness of the resulting minimum values of $R^{(I)}$ can be confirmed by the stability of $F^{(I)}$'s under block transformations.

The flatness of the bottom of these valleys suggests us to restrict the coupling parameter space further to $c_2 = c_3 = 0$. We then get a quite simple action consisting only with plaquettes and rectangles:
$$S_{imp} = \beta \{c_0 \sum \Box + c_1 \sum \Box\!\Box\},$$
with $c_0 = 1 - 8c_1$ and $c_1 = -0.331$ $(-0.293)$ to minimize $R^{(1)}$ $(R^{(2)})$. In the followings we use $c_1 = -0.331$.

A perturbative calculation shows that the scale parameter of this action, $\Lambda_{IM}$, is close to that in the $\overline{MS}$ scheme: $\Lambda_{\overline{MS}}/\Lambda_{IM} \simeq 0.488$ [15] in contrast with the large ratio for the case of the standard action. Therefore, no tadpole improvements [16] are necessary for our action. Correspondingly, $a^{-1} \simeq 1.8 - 1.9$ GeV is obtained at small $\beta = 2.4$ with the improved action, while $\beta \simeq 5.3$ is required with the standard action. This action was applied to numerical studies of the string tension, the hadron spectrum, topological properties and the $U(1)$ problem[17], and the results indicate that $O(a)$ effects with this action are smaller than the standard action. Application to finite temperature gauge theory is in progress.

# 3  Results for $N_F = 2$

As a first step toward an improved full QCD, we combine the RG improved pure gauge action discussed in the previous section with the standard Wilson quark action. This should work well at least at small $K$ where the effects of quarks can be absorbed by renormalizations of pure gauge interactions. Our results for $N_F = 2$ QCD presented below suggests, however, that our action improves the theory very much also near $K_c$. Preliminary results are reported in [9, 4].

Simulations are done on $8^3 \times 4$ and $8^4$ lattices. Fig. 4 shows the result for the phase diagram. The chiral limit $K_c$ is determined by $m_\pi^2 = 0$ on the low temperature $8^4$ lattice. Finite temperature transition line $K_t$ is determined on the $8^3 \times 4$ lattice. Unlike for the standard action (Fig. 1), $K_t$ goes simply away from $K_c$ when we increase $\beta$.

Shown in Fig. 5 is $m_\pi^2$ and $m_q$ as a function of $1/K - 1/K_c$ for various $\beta$. From the same $a^{-1} \simeq 1$ GeV, results at $\beta \simeq 2.0$ should be compared with those at $\beta = 5$ with the standard action shown in Fig. 2. The straight line envelop of $m_\pi^2$ in Fig. 5 corresponds to the chiral behavior $m_\pi^2 \propto m_q$ in the confining phase and deviation from this line signals the transition $K_t$ to the deconfining phase. We note: (1) In contrast with the case of the standard action, $m_q$ shows no phase-dependence. (2) $m_\pi^2$ at $K_c$ decreases smoothly to zero when we decrease $\beta$ toward $\beta_{ct}$. This suggests that the chiral transition is continuous as in the case of the standard action. (3) At $\beta > \beta_{ct}$, $m_\pi^2$ as well as other physical observables are quite smooth at $K_t$. This suggests that the transition there is actually a crossover, in accord with a naive expectation that the second order chiral transition will become a crossover at finite $m_q$.

# 4  Conclusion

Phase structure of finite temperature $N_F = 2$ QCD with Wilson quarks is studied using a RG improved action. As in the case of the standard action, the chiral transition is shown to be continuous. On the other hand, not like the case of the standard action where lattice artifacts make the transition very strong at intermediate $K$, the transition with the improved action becomes quickly smooth when we decrease $K$, in accordance with a naive expectation about the fate of a second order chiral transition at finite $m_q$. Studies for $N_F = 3$ and $2 + 1$ are in progress.

Simulations are performed with HITAC S820/80 at KEK and with QCD-PAX and VPP500/30 at the University of Tsukuba. I would like to thank members of KEK and the members of QCDPAX collaboration for their hospitality and strong support. This work is in part supported by the Grant-in-Aid of Ministry of Education, Science and Culture (No.07NP0401).

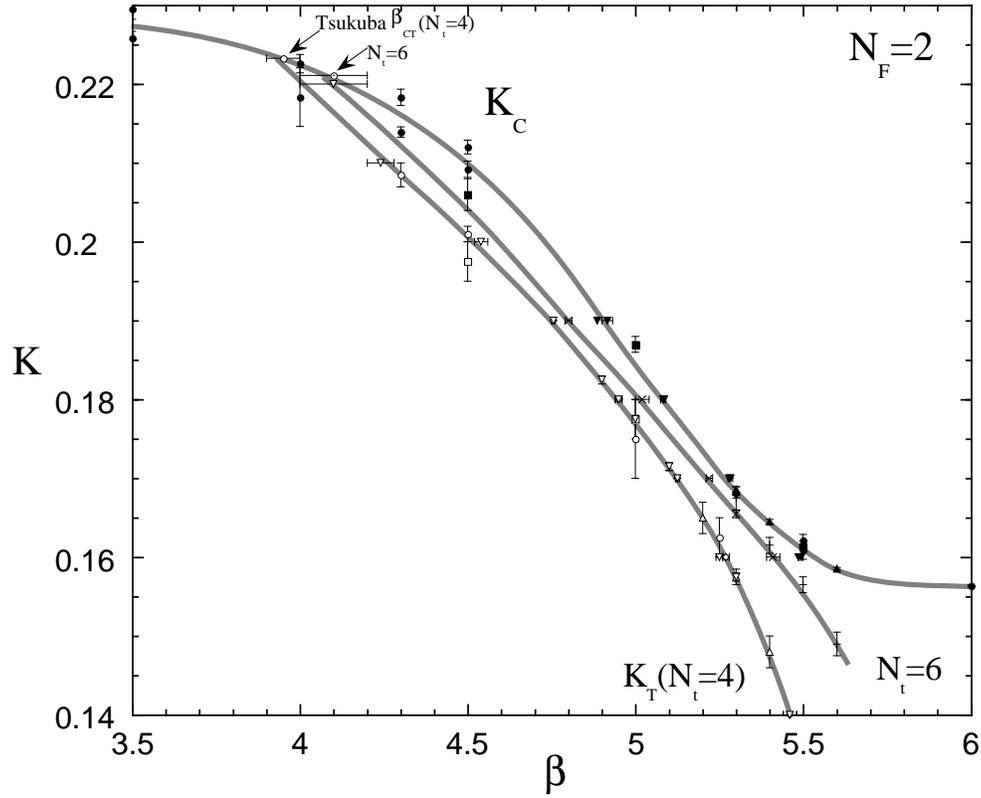

Figure 1: Phase diagram for $N_F = 2$ QCD with Wilson quarks using the standard action. $K_c$ is the chiral limit on $T = 0$ lattices and $K_t(N_t)$'s are for finite temperature transition/crossover. $\beta_{ct}(N_t)$'s are for crossing points of $K_t$ and $K_c$ lines, i.e. the chiral transition. Lines are for guiding eyes.

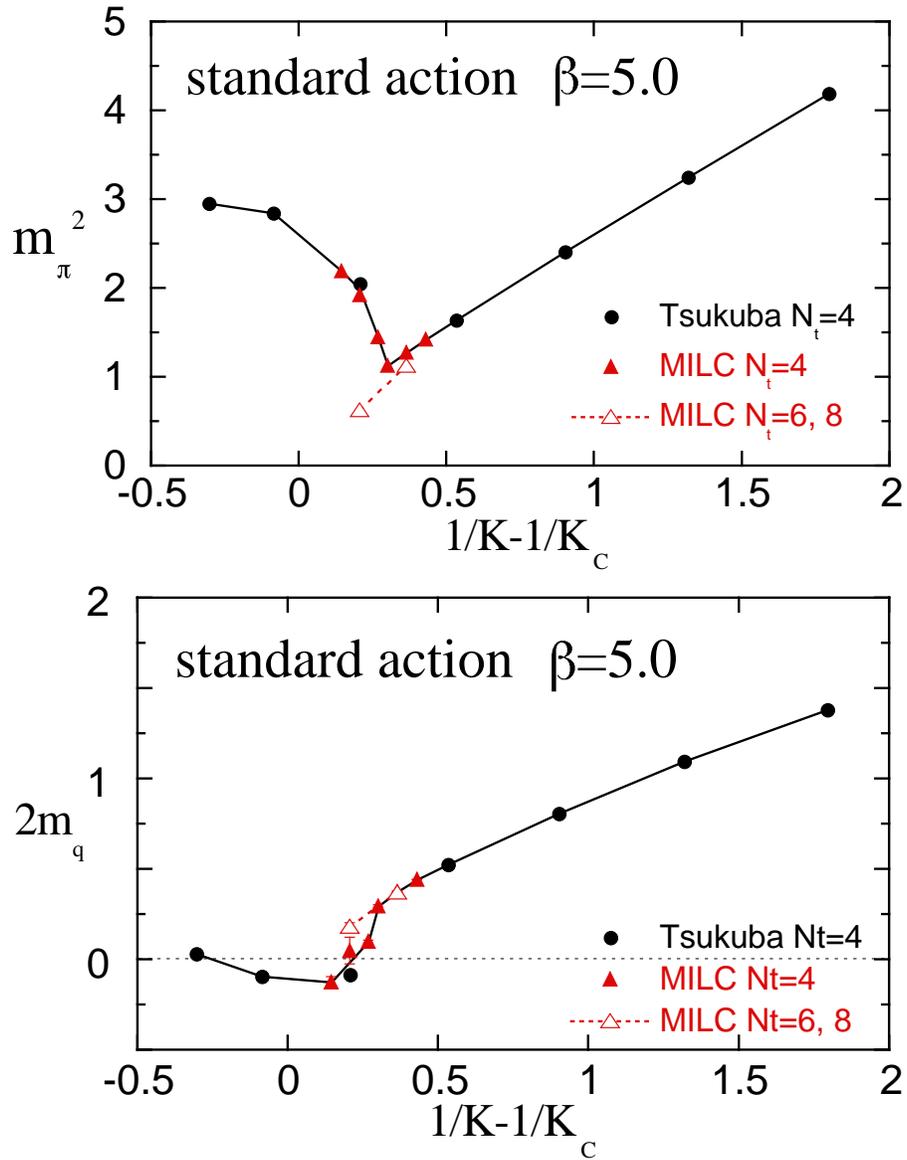

Figure 2: $m_\pi^2$ and $2m_q$ with the standard action at $\beta = 5.0$.

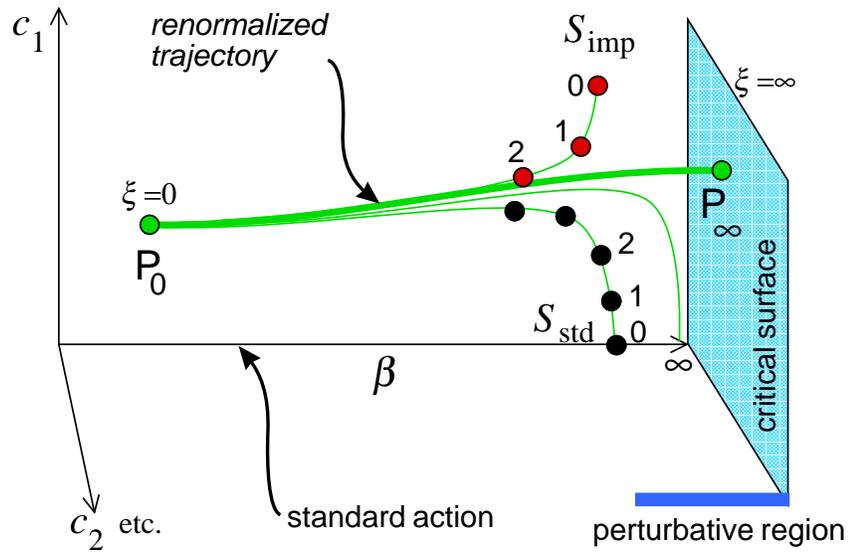

Figure 3: Renormalization group flows in the infinite-dimensional coupling parameter space of QCD.

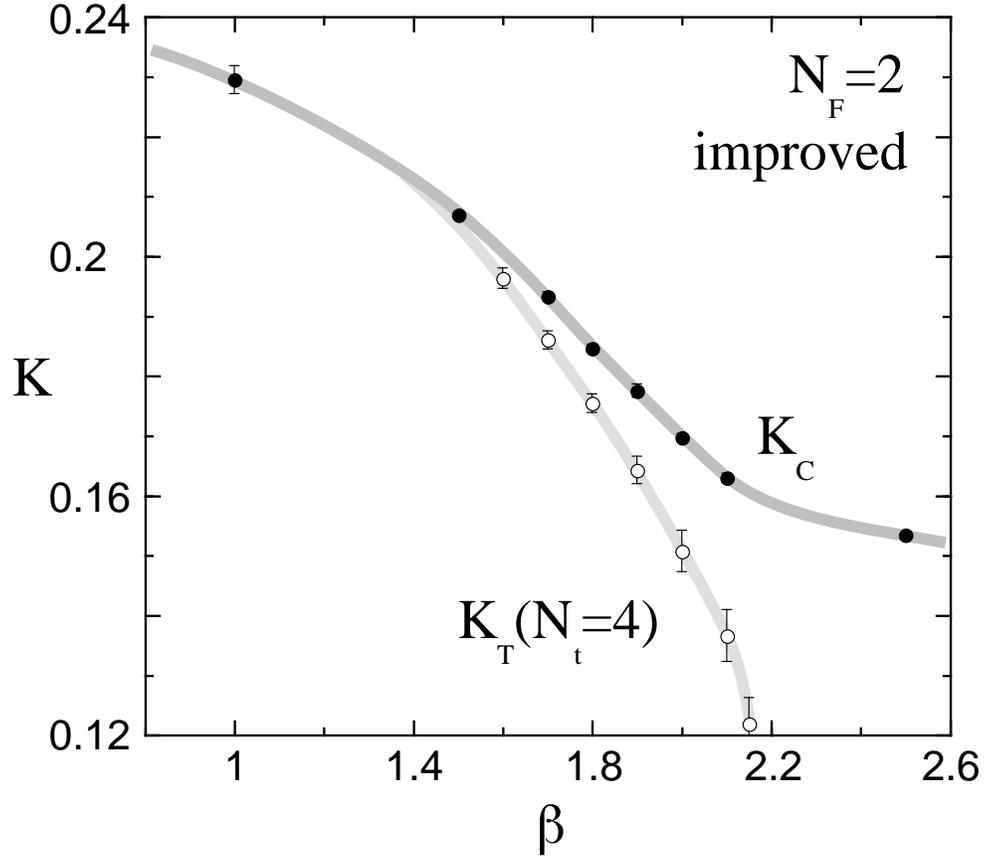

Figure 4: Phase diagram for $N_F = 2$ QCD with Wilson quarks coupled to an RG improved gauge action. $K_c$ is the chiral limit determined by $m_\pi^2 = 0$ on an $8^4$ lattice and $K_t$ is the finite temperature transition/crossover obtained on $8^3 \times 4$. Lines are for guiding eyes.

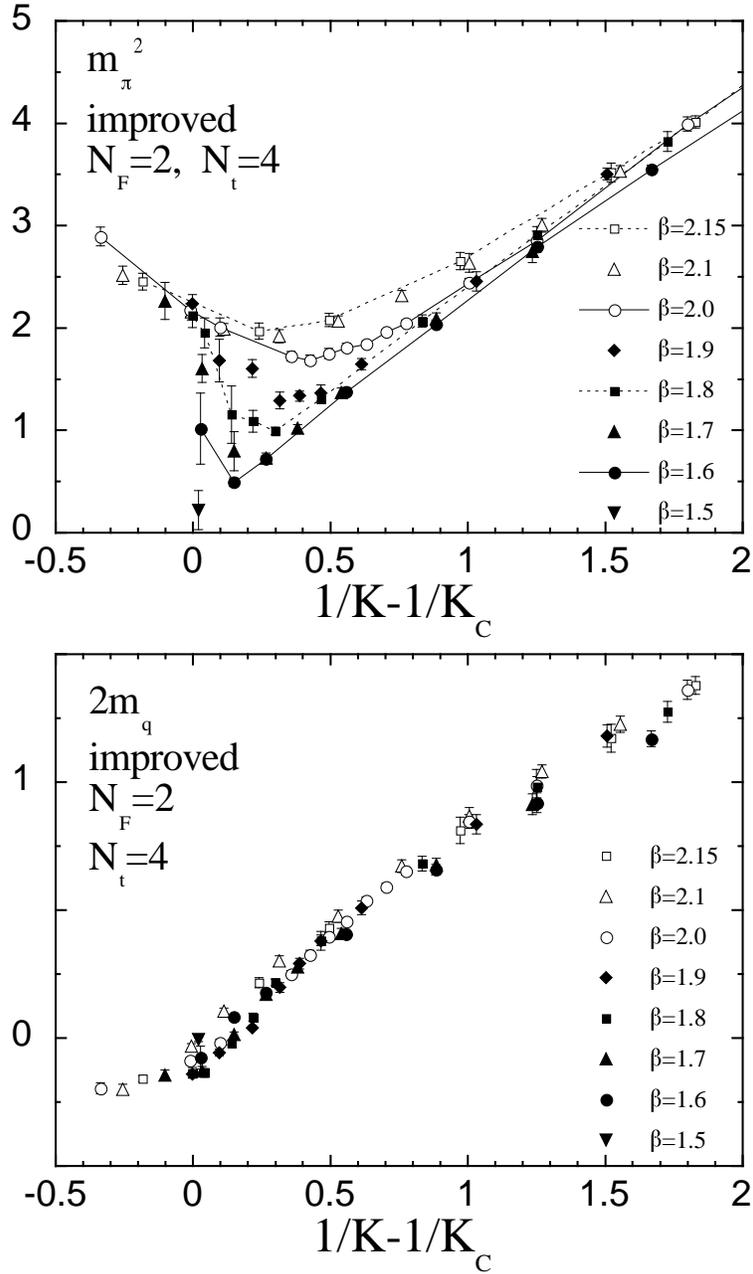

Figure 5: $m_\pi^2$ and $2m_q$ with a RG improved action at various $\beta$.